\documentclass[aps,preprint]{revtex4-2}

\usepackage{graphicx}
\usepackage{bm}
\usepackage{txfonts}
\begin{document}
%\preprint{}

\title{London moment, London's superpotential, Nambu-Goldstone mode, and Berry connection from many-body wave functions}


%\author{Daichi Manabe, Daiki Ishioka}
%\thanks{}
%\affiliation{Graduate school of pure and applied sciences,  University of Tsukuba,Tsukuba, Ibaraki 305-8577, Japan}
%
\author{Hiroyasu Koizumi }
%\thanks{}
\affiliation{Division of Quantum Condensed Matter Physics, Center for Computational Sciences, University of Tsukuba,Tsukuba, Ibaraki 305-8577, Japan}


\date{\today}

\begin{abstract}
Although the standard theory of superconductivity based on the BCS theory is a successful one, there are several experimental results that indicate the necessity for fundamental revisions. One of them is the mass in the London moment. Experiments indicate the mass in the London moment is the free electron mass although the BCS theory and its extension predict it to be an effective mass.

We show that this discrepancy is lifted if we install the London's superpotential in the theory, and identify it as the Berry phase arising from the many-body wave functions. Then, the induced current by the applied magnetic field becomes a stable current calculated
using the free energy in contrast to the linear response current assumed in the standard theory which yields the Nambu-Goldstone mode. The Nambu-Goldstone mode arising from the breakdown of the global $U(1)$ gauge invariance in the standard theory is replaced by the collective mode arising from the Berry connection. Then, the free electron mass appears in the London moment. 

\end{abstract}

%\keywords{}


\maketitle

\section{Introduction}

The discovery of the Meissner effect \cite{Meissner} has made it clear that the normal-superconducting phase transition in a magnetic field in the $H$-$T$ plane (for Type I superconductors) is reversible. 
Also it is notable that a series of work \cite{Keesom1934a,Keesom1934b,Keesom1937,Keesom} indicate that the superconducting-normal state transition in the presence of a magnetic field occurs without energy dissipation, and the state of the art calorimetry indicates that 99.99\% of the supercurrent stops without current carriers undergoing irreversible collisions (see Appendix B of Ref.~\cite{Hirsch2017}). 
Thus, it is legitimate to study superconductivity using thermodynamics. Indeed, the phenomenological Ginzburg-Landau theory that utilizes a free-energy has been successfully applied to various superconducting phenomena \cite{GL}.

 According to the standard theory based on the BCS one \cite{BCS1957}, paired electrons flow without dissipation but single electrons flow with dissipation. Thus, the supercurrent generated by the flow of electron pairs in the magnetic field inevitably produces the Joule heat during the superconducting to normal phase transition due to the existence of a significant number of broken pairs that flow with dissipation \cite{Hirsch2017,Hirsch2018,Hirsch2020}, which contradicts the reversible normal-superconducting phase transition in a magnetic field.
This indicates serious revisions are needed in the standard theory and it has been claimed that the supercurrent should be considered as a collection of loop currents produced by the non-trivial Berry connection \cite{koizumi2020b}.

In the present work, we take up a related discrepancy in the standard theory. 
It concerns with the mass of the {\em London moment} \cite{London1950}.
The London moment is the magnetic field ${\bf B}^{\rm em}$ produced inside rotating superconductors.
 Let ${\bm \omega}$ be the angular frequency of the rotation, the magnetic field ${\bf B}^{\rm rm}$ appears in the
 superconductor is given by
\begin{eqnarray}
{\bf B}^{\rm em}=-{{2m c} \over q} {\bm \omega}
\label{LondonMoment}
\end{eqnarray}
where $m$ is mass, $q$ is charge, and $c$ is the speed of light.
The London moment has been measured many times using different materials, ranging from the conventional superconductor \cite{Hildebrandt1964,Zimmerman1965,Brickman1969,Tate1989,Tate1990} to the high T$_{\rm c}$ cuprates \cite{VERHEIJEN1990a,Verheijen1990} and
heavy fermion superconductors \cite{Sanzari1996}. The results always indicate that the mass $m$ is the free electron mass $m_e$, not the effective mass $m^{\ast}$ predicted by the standard theory.

The appearance of the magnetic field in Eq.~(\ref{LondonMoment}) was first realized by Becker et al. \cite{Becker1933}.
It arises from the supercurrent flowing in the surface region of the rotating superconductor.
In the standard theory this supercurrent is a flow of the Cooper pairs with the effective mass of two electrons $2m^{\ast}$ with the charge $q=-2e$, where $-e$ is the electron charge.
If we take the view that the supercurrent is a dissipationless flow of paired electrons and use $q=-2e$, we should have $m=2m^{\ast}$, which contradicts the experimental value $m=2m_e$ for this case (actually, we could put $q=-e$ and $m=m_e$). The main concern of the present work is to present the reason for the appearance of this mass in the London moment.

Let us examine the part of the London's theory of superconductivity that explains the appearance of the London moment and its implication \cite{London1950}.
London argues that the following relation (the {\em London equations}) holds in superconductors
\begin{eqnarray}
\partial_{\ell} u_k -\partial_k u_{\ell}=-{ q \over {mc}} f_{\ell k}
\label{Londoneq}
\end{eqnarray}
where $u_k$ is the four velocity vector, and $f_{\ell k}$ is the electromagnetic field tensor given by
\begin{eqnarray}
 f_{\ell k}=\partial_{\ell} A_k -\partial_k A_{\ell}
 \end{eqnarray}
where $A_k$ is the four electromagnetic vector potential.

The spatial part of Eq.~(\ref{Londoneq}) is given by
\begin{eqnarray}
\nabla \times {\bf v}=-{ q \over {mc}} {\bf B}^{\rm em}
\label{Londoneq2}
\end{eqnarray}
where ${\bf v}$ is the velocity vector.

Actually, the original London equations are those expressed using the current density \cite{London1935},
\begin{eqnarray}
\nabla \times \Lambda {\bf j}=-{c} {\bf B}^{\rm em}, \quad \Lambda ={{nq^2} \over m}
\label{Londoneq3}
\end{eqnarray}
where $n$ is the density of the charge carriers.
It is well-known that this explains the Meissner effect when it is combined with the one of Maxwell's equations
\begin{eqnarray}
\nabla \times {\bf B}^{\rm em}={ {4 \pi} \over c} {\bf j}
\end{eqnarray}

However, the use of the new ones given in Eq.~(\ref{Londoneq}) using the velocity ${\bf v}$ is crucial to obtain the correct mass for the London moment as will be shown below.

Let us consider a superconducting sphere with its center at the origin and rotate it about its symmetry axis with constant angular velocity ${\bm \omega}$.
Then, the velocity at ${\bf r}$ in the sphere is given by
\begin{eqnarray}
{\bf v}={\bm \omega} \times {\bf r}
\end{eqnarray}
Electrons inside the superconductor will move with this velocity to shield the background positive charge.
Substituting this in Eq.~(\ref{Londoneq2}) immediately yields Eq.~(\ref{LondonMoment}).

London argues that behind Eq.~(\ref{Londoneq}) is a scalar potential $\chi^{\rm super}$.
Let us consider the four momentum vector $p_k$ given by
\begin{eqnarray}
p_k=m u_k+{ q \over c}A_k
\label{Londoneq3}
\end{eqnarray}
and expressed the equation in Eq.~(\ref{Londoneq}) as
\begin{eqnarray}
\partial_{\ell} p_k -\partial_k p_{\ell}=0
\end{eqnarray}

This relation is satisfied if a scalar $\chi^{\rm super}$ called the {\em superpotential} exists in superconductors,
and $p_k$ is given by
\begin{eqnarray}
p_k=\hbar \partial_k \chi^{\rm super}
\end{eqnarray}

In the following we use the spatial part of Eq.~(\ref{Londoneq3}) given by
\begin{eqnarray}
\hbar \nabla \chi^{\rm super}=m {\bf v}+{ q \over c}{\bf A}^{\rm em}
\label{Londone4}
\end{eqnarray}
where ${\bf A}^{\rm em}$ is the electromagnetic vector potential.

London further argues that the wave function for the superconducting state is rigid in the sense that
when a magnetic field is applied, the wave function is modified only by the change of the phase factor
\begin{eqnarray}
\Psi_0 \rightarrow \Psi=\Psi_0 e^{{ i } \sum_j \chi^{\rm super} ({\bf r}_j)}
\label{LondonWF}
\end{eqnarray}
where $\Psi_0$ is the wave function when magnetic field is absent, and  $\Psi$ is that for the case with a magnetic field.

As is shown in the above argument, the London theory for superconductivity is composed of the superpotential $\chi^{\rm super}$ and rigid wave function $\Psi_0$.
The BCS theory successfully explains how the rigidity of the superconducting wave function is realized. Does the standard theory have something corresponding to the superpotential?
We may identify the Nambu-Goldstone mode \cite{Nambu1960} as something corresponding to the superpotential potential $\chi^{\rm super}$.
However, if this is the case, the mass $m$ in Eq.~(\ref{LondonMoment}) will be the effect mass $m^{\ast}$.
Thus, we need to find a different object for $\chi^{\rm super}$.

If we rewrite the relation in Eq.~(\ref{Londone4}) as
\begin{eqnarray}
m {\bf v}=-{ q \over c} \left( {\bf A}^{\rm em}  - { {c \hbar} \over q} \nabla \chi^{\rm super} \right)
\label{eqr1}
\end{eqnarray}
we notice that $m{\bf v}$ is gauge invariant, thus, ${\bf A}^{\rm eff}$ given by
\begin{eqnarray}
{\bf A}^{\rm eff} = {\bf A}^{\rm em}  - { {c \hbar} \over q} \nabla \chi^{\rm super}
\label{eqr2}
\end{eqnarray}
must be gauge invariant. 

It has been argued that the fictitious magnetic field ${\bf A}^{\rm fic}$ that makes
\begin{eqnarray}
{\bf A}^{\rm eff} = {\bf A}^{\rm em} + {\bf A}^{\rm fic}
\label{eqr3}
\end{eqnarray}
gauge invariant can arise as a Berry connection \cite{HKoizumi2013,koizumi2020,koizumi2020c}. 
In the present work, we identify this ${\bf A}^{\rm fic}$ as $- { {c \hbar} \over q} \nabla \chi^{\rm super}$.
In this case, the supercurrent generation is attributed to the appearance of non-trivial connection of geometry in mathematics.
Then, the mass should not be the effective mass defined as the inertial mass in the solid when a force is exerted, but something for
the geometric object.
Actually, we will show that in this case the mass $m=m_e$ appears in the London moment.

We also concern the relation between the Nambu-Goldstone mode and the superpotential in this work.
The Nambu-Goldstone mode in the standard theory appears to restore the gauge invariance in the induced supercurrent by a magnetic field.
Since thermodynamics can be applied to superconducting phenomena, the supercurrent can be calculated using the free energy $F$ that is a functional of ${\bf A}^{\rm em}$ \cite{Schafroth}. The relations Eqs.~(\ref{eqr1}), (\ref{eqr2}), and (\ref{eqr3})
indicate that ${\bf A}^{\rm em}$ always appears as a part of ${\bf A}^{\rm eff}$. Thus, the supercurrent is given by
\begin{eqnarray}
{\bf j}({\bf r})=-c {{\delta F[{\bf A}^{\rm em} + {\bf A}^{\rm fic}] } \over {\delta {\bf A}^{\rm em}({\bf r})} }=-c {{\delta F[{\bf A}^{\rm em} + {\bf A}^{\rm fic}] } \over {\delta {\bf A}^{\rm fic}({\bf r})} }
\label{eqMeisnner1}
\end{eqnarray}

In the standard theory, the induced current by the magnetic field is calculated using the linear response theory. 
We may regard this current as the linear approximation to the above given by
\begin{eqnarray}
{ j}_\mu ({\bf r}) \approx -c \sum_\nu \int d^3 r'  { {\delta^2 F[{\bf A}^{\rm em} + {\bf A}^{\rm fic}] } \over { \delta {A}_{\mu}^{\rm em}({\bf r}) \delta {A}_{\nu}^{\rm em}({\bf r}')} }  {A}_{\nu}^{\rm eff}
\label{eqMeisnner2}
\end{eqnarray}
From this view, we will argue that the Nambu-Goldstone mode is replaced by the collective mode arising from the Berry connection.

 %In the present work, we argue that the superpotential of London can be identified as the Berry phase arising from the many-body wave function \cite{koizumi2019}, and $\nabla \chi^{\rm super}$ to its connection. The Berry phase was not known during the time of development of the standard theory \cite{Berry}.
 %It has been argued that it has to be added as a necessary ingredient of superfluidity and superconductivity \cite{koizumi2019}.
%By attributing $\nabla \chi^{\rm super}$ to the Berry connection, we will show that the standard theory can be made to the one that explains the London magnetic moment with free electron mass.
 
 The organization of the present work  is following;
 in Section~\ref{sec2}, we explain the Berry connection for many-body wave functions. In Section~\ref{sec3}, the collective mode arising from the Berry connection and associated number changing operators are explained. The relation between the standard theory and the present one using the Berry connection is given in Section~\ref{sec4}; here, the mass in the London moment is shown to be the free electron mass.
 The relation between the Ginzburg-Landau theory and the present theory is given in Section~\ref{sec5}.
 The relation between the collective mode arising from the Berry connection and the Nambu-Goldstone mode is explored in Section~\ref{sec6}.
 Lastly, we conclude the present work in Section~\ref{sec7}.

\section{Berry connection from many-body wave functions}
\label{sec2}

Let us first briefly review the Berry phase and Berry connection by following the Berry's derivation \cite{Berry}.

We consider slow dynamical variables ${\bf R}$ whose momentum operator is $-i\hbar \nabla_{\bf R}$,
and fast ones ${\bf r}$ whose momentum operator $-i\hbar \nabla_{\bf r}$.

The Hamiltonian for the ${\bf r}$ dynamics, which we denote as $H({\bf R})$, depends on ${\bf R}$; we first solve it by treating ${\bf R}$ as parameters. An example of such a treatment is the Born-Oppenheimer approximation \cite{BornOppenheimer}, where $H({\bf R})$ is the Hamiltonian for the electronic problem whose dynamical variables are electron coordinates; the slow dynamical variables are the nuclear coordinates.

The equation for the fast variable problem is the following Schr\"{o}dinger equation 
\begin{eqnarray}
H({\bf R}) |n({\bf R}) \rangle=  E_n({\bf R})|n({\bf R}) \rangle 
\end{eqnarray}

Let us consider the following time-dependent Schr\"{o}dinger equation corresponding to the above
\begin{eqnarray}
i \hbar \partial _t |\psi_n \rangle =H({\bf R}(t)) |\psi_n\rangle 
\end{eqnarray}
with including the change of ${\bf R}$ in time.

We consider the situation where the fast variable state adheres to a single state $|n({\bf R}(t)) \rangle$, and write the time-dependent wave function as
\begin{eqnarray}
|\psi_n \rangle= e^{- { i\over \hbar} \int_0^t dt'E_n({\bf R}(t')) } e^{ i\gamma_n (t)} |n({\bf R}(t)) \rangle
\end{eqnarray}
This solution assumes that the state is in $|n({\bf R}(t)) \rangle$ with a possible phase factor change $e^{ i\gamma_n (t)}$.

The phase $\gamma_n$ is called the {\em Berry phase}. A simple calculation yields 
\begin{eqnarray}
\gamma_n(t)=-\int_{{\bf R}(0)}^{{\bf R}(t)} d{\bf R}\cdot {\bf A}_n({\bf R})
\end{eqnarray}
where ${\bf A}_n({\bf R})$ is the {\em Berry connection}
\begin{eqnarray}
{\bf A}_n({\bf R})=-i\langle n({\bf R}) |\nabla_{\bf R} |n({\bf R}) \rangle 
\end{eqnarray}

Now consider the effect of the Berry phase on the dynamics for the slow variables ${\bf R}$. 
Since the fast variable state adheres to $|n({\bf R}) \rangle$, we express 
the total wave function as $f({\bf R}) \langle {\bf r}|n({\bf R}) \rangle$.

Application of the momentum operator for the ${\bf R}$ motion on the total wave function gives
\begin{eqnarray}
 -i\hbar \nabla_{\bf R} f({\bf R})  \langle {\bf r}|n({\bf R}) \rangle = \langle {\bf r}|n({\bf R}) \rangle[-i\hbar \nabla_{\bf R}f({\bf R}) ]
 - f({\bf R}) [i\hbar \nabla_{\bf R} \langle {\bf r}|n({\bf R}) \rangle]
 \label{eqgauge}
 \end{eqnarray}
 
 Since the fast variable state stays in $|n({\bf R}) \rangle$, the effective one on $f({\bf R})$ is obtained by 
 multiplying $\langle n({\bf R})|{\bf r} \rangle$ on the left of Eq.~(\ref{eqgauge}) and integrating over ${\bf r}$,
\begin{eqnarray}
\langle n ({\bf R})|n({\bf R}) \rangle[-i\hbar \nabla_{\bf R}f({\bf R}) ]
 - f({\bf R}) i\hbar \langle n ({\bf R})|\nabla_{\bf R} |n({\bf R}) \rangle=[-i\hbar \nabla_{\bf R} f({\bf R}) +\hbar {\bf A}_n({\bf R})]f({\bf R})
\end{eqnarray}

This indicates that the ${\bf R}$ dynamics is the one in the gauge field $\hbar {\bf A}_n$, and the following replacement takes place 
\begin{eqnarray}
-i \hbar \nabla_{\bf R}  \rightarrow  -i\hbar \nabla_{\bf R}+\hbar {\bf A}_n  
\end{eqnarray}
 as in the system in an electromagnetic field.
We may view this change as the appearance of the connection of geometry in mathematics. 
This connection may be non-trivial and cause important effects \cite{Mead79,Yuan1289}.

We apply the above argument to a $N$ particle system with the wave function 
\begin{eqnarray}
\Phi ({\bf r}_1, \cdots, {\bf r}_{N},t)
\end{eqnarray}
where ${\bf r}_j$ is the coordinate of the $j$th particle.
We consider the appearance of a Berry connection in the velocity field for ${\bf r}_1$ through 
the interaction with the other dynamical variables, ${\bf r}_2, \cdots, {\bf r}_{N}$.
In this case, we cannot say that ${\bf r}_1$ is a slow variable and the others are fast ones.
Nevertheless, we can calculate the Berry connection in the following manner.

First, we define the parameterized wave function $|n_{\Phi}({\bf r}_1) \rangle$ with the parameter ${\bf r}_1$, 
 \begin{eqnarray}
\langle  {\bf r}_{2}, \cdots, {\bf r}_{N} |n_{\Phi}({\bf r}_1,t) \rangle = { {\Phi({\bf r}_1, {\bf r}_{2}, \cdots, {\bf r}_{N},t)} \over {|C_{\Phi}({\bf r}_1 ,t)|^{{1 \over 2}}}}
\end{eqnarray}
where $|C_{\Phi}({\bf r}_1,t)|$ is the normalization constant given by 
\begin{eqnarray}
|C_{\Phi}({\bf r}_1,t)|=\int d{\bf r}_{2} \cdots d{\bf r}_{N}\Phi({\bf r}_1, {\bf r}_{2}, \cdots)\Phi^{\ast}({\bf r}_1,  {\bf r}_{2}, \cdots)
\end{eqnarray}
$|n_{\Phi}({\bf r}_1) \rangle$ is normalized to one, and it is a single-valued function of the parameter ${\bf r}_1$.

Using $|n_{\Phi}\rangle$, the {\em Berry connection for many-body wave function} is calculated as
 \begin{eqnarray}
{\bf A}^{\rm MB}_{\Phi}({\bf r}_1,t)=-i \langle n_{\Phi}({\bf r}_1,t) |\nabla_{{\bf r}_1} |n_{\Phi}({\bf r}_1,t) \rangle
\label{eqBerryConnection}
\end{eqnarray}

The appearance of ${\bf A}^{\rm MB}_{\Phi}({\bf r}_1,t)$ will cause the following modification 
\begin{eqnarray}
-i \hbar \nabla_{{\bf r}_1}  \rightarrow  -i\hbar \nabla_{{\bf r}_1}+\hbar {\bf A}^{\rm MB}_{\Phi}({\bf r}_1)
\label{eqFBerry}
\end{eqnarray}
to the dynamics for the variable ${\bf r}_1$ if the system stays in $\Phi$.
 The same change also occurs for the other variables, ${\bf r}_2, \cdots, {\bf r}_{N}$.
 
The connection ${\bf A}^{\rm MB}_{\Phi}({\bf r}_1,t)$
arises through the wave function that also depends on the other variables, ${\bf r}_2, \cdots, {\bf r}_N$.
The dynamics stays on the same wave function in the sense that specifying the value for ${\bf r }_1$
determines the wave function for the dynamical variables ${\bf r}_2, \cdots, {\bf r}_N$, uniquely, because only a single wave function
is considered. This fact replaces the adiabaticity requirement assumed in the original derivation of the Berry phase.

In order to make the existence of  ${\bf A}^{\rm MB}_{\Phi}$ apparent
we express $\Phi ({\bf r}_1, \cdots, {\bf r}_{N},t)$ as
 \begin{eqnarray}
\Phi =\Phi_0 \exp\left(i \sum_{j=1}^{N} \int_{0}^{{\bf r}_j} {\bf A}_{\Phi}^{\rm MB}({\bf r}',t) \cdot d{\bf r}' \right)
\label{Phi}
\end{eqnarray}

In the non-relativistic approximation, the kinetic energy part of the Hamiltonian is given by
\begin{eqnarray}
K_0={ 1\over {2m}} \sum_{j=1}^{N_b} \left( {\hbar \over i} \nabla_{j} \right)^2
\label{a2}
\end{eqnarray}
where $m$ is the particle mass. $m=m_e$ for the electron system.
We can see that for the wave functionn$\Phi_0 $, the kinetic energy becomes the one with 
the replacement $-i \hbar \nabla_{\bf r}  \rightarrow  -i\hbar \nabla_{\bf r}+\hbar {\bf A}^{\rm MB}_{\Phi}$.

We can express $\Phi_0$ as
\begin{eqnarray}
\Phi_0 ({\bf r}_1, \cdots, {\bf r}_{N_b},t)=\Phi ({\bf r}_1, \cdots, {\bf r}_{N},t)\exp\left(- i \sum_{j=1}^{N} \int_{0}^{{\bf r}_j} {\bf A}_{\Phi}^{\rm MB}({\bf r}',t) \cdot d{\bf r}' \right)
\end{eqnarray}
Then, it can be shown that this is a currentless state for the current operator associated with $K_0$.

Now, we assume that the origin of ${\bf A}_{\Phi}^{\rm MB}$ is not the ordinary electromagnetic field, and satisfies
\begin{eqnarray}
\nabla \times {\bf A}^{\rm MB}_{\Phi}=0
\end{eqnarray}

Then, we may write 
\begin{eqnarray}
 {\bf A}^{\rm MB}_{\Phi}=\nabla \varphi
 \label{varphi1}
\end{eqnarray}
using a scalar function. Note that $\varphi$ may be a multi-valued function of the coordinate.

Now the many-body wave function is given by
 \begin{eqnarray}
\Phi =\Phi_0 \exp\left(i \sum_{j=1}^{N} \varphi({\bf r},t) \right)
 \label{varphi2}
\end{eqnarray}
according to Eqs.~(\ref{Phi}) and (\ref{varphi1}).

This wave function has the form for the superconducting state envisaged by London in Eq.~(\ref{LondonWF}).
The phase variable $\varphi$ may play the role of the superpotential $\chi^{\rm super}$.
When the rigidity of the wave function is realized in $\Phi_0$, the low energy physics will be described by the collective mode associated with $\varphi$.

\section{Collective mode $\varphi$ and associated number changing operators}
\label{sec3}

Let us treat $\varphi$ in Eqs.~(\ref{varphi2}) as a collective dynamical variable. 
 For this purpose, we use the time-dependent variational principle using the following Lagrangian,
\begin{eqnarray}
{\cal L}\!=\langle \Phi | i\hbar \partial_t \!-\!H| \Phi \rangle\!=\! i\hbar \langle \Phi_0 | \partial_t | \Phi_0 \rangle- \hbar \int \!d{\bf r} \ {{\rho \dot{\varphi} }} - \langle \Phi |H| \Phi \rangle
\label{L}
\end{eqnarray}
where $\rho$ is the number density of the particles \cite{Koonin1976}.

From the above Lagrangian, the conjugate momentum of $\varphi$ is obtained as
\begin{eqnarray}
p_{\varphi}= {{\delta {\cal L}} \over {\delta \dot{\varphi}}}=-\hbar \rho
\label{momentumchi}
\end{eqnarray}
thus, $\varphi$ and $\rho$ are canonical conjugate variables.

If we follow the canonical quantization condition 
\begin{eqnarray}
[\hat{p}_{\varphi}({\bf r}, t), \hat{\varphi}({\bf r}', t)]=-i\hbar \delta ({\bf r}- {\bf r}')
\end{eqnarray}
where $\hat{p}_{\varphi}$ and $\hat{\varphi}$ are operators corresponding to ${p}_{\varphi}$ and ${\varphi}$ respectively, 
we obtain the following relation
\begin{eqnarray}
\left[{ {\hat{\rho}({\bf r}, t)} } , \hat{\varphi}({\bf r}', t) \right]=i \delta ({\bf r}- {\bf r}')
\label{commu0}
\end{eqnarray}
where $\hat{\rho}$ is the operator corresponding to $\rho$.

 Strictly speaking, $\hat{\varphi}$ is not a hermitian operator; however, it is known that when it is used as 
 $\sin \hat{\varphi}$ or  $\cos \hat{\varphi}$, the problem is avoided, practically \cite{Phase-Angle}. In the following we use
 $\hat{\varphi}$  as $e^{\pm i {\hat{\varphi} }}$.
 
We construct the following boson field operators from $\hat{\varphi}$ and $\hat{\rho}$,
\begin{eqnarray}
\hat{\psi}^{\dagger}({\bf r})= \left(\hat{\rho}({\bf r}) \right)^{1/2} e^{- i {\hat{\varphi}({\bf r}) }}, \quad \hat{\psi}({\bf r})= e^{i { \hat{\varphi}({\bf r})}}\left( \hat{\rho}({\bf r}) \right)^{1/2}
\label{boson1}
\end{eqnarray}

Using Eq.~(\ref{commu0}), the following relations are obtained,
\begin{eqnarray}
 [\hat{\psi}({\bf r}),\hat{\psi}^{\dagger}({\bf r}')]=\delta({\bf r}-{\bf r}'), \quad  [\hat{\psi}({\bf r}),\hat{\psi}({\bf r}')]=0, \quad  [\hat{\psi}^{\dagger}({\bf r}),\hat{\psi}^{\dagger}({\bf r}')]=0
 \label{commu2}
\end{eqnarray}
  
From Eqs.~(\ref{commu0}) and (\ref{commu2}), the following relations are obtained 
   \begin{eqnarray}
[e^{\pm i  \hat{\varphi}({\bf r})}, \hat{\psi}^{\dagger}({\bf r}')\hat{\psi}({\bf r}')]=\pm e^{ \pm i  \hat{\varphi}({\bf r})}\delta({\bf r}-{\bf r}')
\label{commchi}
 \end{eqnarray}
 
Now we define the state vector $| \rho({\bf r})\rangle$ that satisfies
     \begin{eqnarray}
  \hat{\psi}^{\dagger}({\bf r}')\hat{\psi}({\bf r}')| \rho({\bf r})\rangle = \rho({\bf r}')| \rho({\bf r})\rangle
 \end{eqnarray}
 
 Then, $e^{ \pm i  \hat{\varphi}({\bf r}')}$ are number changing operators that satisfies
    \begin{eqnarray}
  e^{ \pm i  \hat{\varphi}({\bf r}')}| \rho({\bf r})\rangle  \propto| \rho({\bf r}) \mp \delta({\bf r}-{\bf r}')\rangle,
\label{commchi2}
 \end{eqnarray}
indicating $e^{  i  \hat{\varphi}({\bf r}')}$ reduces and  $e^{ -i  \hat{\varphi}({\bf r}')}$ increases the number of particles by one at ${\bf r}'$.
We will define the particle number conserving Bogoliubov operators in the following.

 \section{Relation to the BCS theory}
 \label{sec4}
 
First, we show that the number changing operators $e^{ \pm i  \hat{\varphi}({\bf r})}$ considered in the previous section are `hiding' as phase factors in the standard theory.

The BCS superconducting state vector is given by
\begin{eqnarray}
|{\rm BCS} (\theta) \rangle=\prod_{\bf k}\left(u_{\bf k}+v_{\bf k}c^{\dagger}_{{\bf k} \uparrow}c^{\dagger}_{-{\bf k} \downarrow}
e^{ {i}{\theta}} \right)|{\rm vac} \rangle
\label{BCS}
\end{eqnarray}
where real parameters $u_{\bf k}$ and $v_{\bf k}$ satisfy $u_{\bf k}^2+v_{\bf k}^2=1$.
It is a linear combination of different particle number states, thus, breaks the global $U(1)$ gauge invariance. 

Now we define the Bogoliubov operators \cite{Bogoliubov58} given by
\begin{eqnarray}
 \gamma_{{\bf k} \uparrow }&=& u_{k} e^{-{i \over 2}\theta} c_{{\bf k} \uparrow} -  v_{ k} e^{{i \over 2}\theta} c^{\dagger}_{-{\bf k} \downarrow}
\nonumber
\\
\gamma_{-{\bf k} \downarrow }&=& u_{k} e^{-{i \over 2}\theta} c_{-{\bf k} \downarrow} +  v_{ k} e^{{i \over 2}\theta} c^{\dagger}_{{\bf k} \uparrow}
\end{eqnarray}

Using the above operators, the superconducting ground state is defined as the ground state for the Bogoliubov excitations,
\begin{eqnarray}
 \gamma_{{\bf k} \uparrow }|{\rm BCS} \rangle=0, \quad \gamma_{-{\bf k} \downarrow }|{\rm BCS} \rangle=0
 \end{eqnarray}
 We will use the above as the definition of the BCS ground state rather than the one given in Eq.~(\ref{BCS}).
 The Bogoliubov excitations are equipped with an energy gap from the electron pairing, and realizes the rigidity in the wave function.
 
 In the original BCS theory, the normal metallic state is assumed to be well-described by the free electrons with the effective mass $m^{\ast}$.
 Then, the electron field operators are given by
\begin{eqnarray}
\hat{\Psi}_{\sigma}({\bf r})={ 1 \over \sqrt{V}}\sum_{\bf k} e^{i {\bf k} \cdot {\bf r}} c_{{\bf k} \sigma}
\end{eqnarray}

In the superconducting state, the Bogoliubov operators are more appropriate. Thus, we rewrite the field operators as
\begin{eqnarray}
\hat{\Psi}_{\uparrow}({\bf r})&=&{ 1 \over \sqrt{V}}\sum_{\bf k} e^{{i \over 2}\theta}e^{i {\bf k} \cdot {\bf r}} \left( \gamma_{{\bf k} \uparrow } u_{ k} -  \gamma^{\dagger}_{-{\bf k} \downarrow } v_{k} \right)
\nonumber
\\
\hat{\Psi}_{\downarrow}({\bf r})&=&{ 1 \over \sqrt{V}}\sum_{\bf k} e^{{i \over 2}\theta}e^{i {\bf k} \cdot {\bf r}} \left( \gamma_{{\bf k} \downarrow } u_{ k} +\gamma^{\dagger}_{-{\bf k} \uparrow } v_{ k} \right)
\end{eqnarray}

The transformation from the standard theory to the present theory starts with replacing $e^{\pm{i \over 2}\theta}$ by the number changing operators $e^{\pm {i \over 2}\hat{\varphi}}$. Here, we do not specify the coordinate in $\hat{\varphi}$ (we will introduce the coordinate dependence, later). Then, the Bogoliubov transformation becomes
\begin{eqnarray}
 \gamma_{{\bf k} \uparrow }&=& u_{k} e^{-{i \over 2}\hat{\varphi}} c_{{\bf k} \uparrow} -  v_{ k} e^{{i \over 2}\hat{\varphi}} c^{\dagger}_{-{\bf k} \downarrow}
\nonumber
\\
\gamma_{-{\bf k} \downarrow }&=& u_{k} e^{-{i \over 2}\hat{\varphi}} c_{-{\bf k} \downarrow} +  v_{ k} e^{{i \over 2}\hat{\varphi}} c^{\dagger}_{{\bf k} \uparrow}
\end{eqnarray}

The important point here is that the above Bogoliubov operators conserve particle numbers.
Terms like $e^{-{i \over 2}\hat{\varphi}} c_{{\bf k} \sigma}$ can be interpreted that an electron in the $({\bf k}, \sigma)$ single-electron mode is 
annihilated and an electron is added to the collective mode described by $\varphi$; those like $e^{{i \over 2}\hat{\varphi}} c^{\dagger}_{{\bf k} \sigma}$ create an electron in the $({\bf k}, \sigma)$ single-electron mode and subtract an electron from the collective mode described by $\varphi$. Thus, the Bogoliubov operators cause the fluctuation of the number of electrons in the collective mode. The ground state is the one with fluctuating number of electrons in the collective mode \cite{koizumi2019}.
This state replaces the state with fluctuating total number of electrons in the standard theory.

By including ${1 \over \sqrt{V}}e^{ i {\bf k} \cdot {\bf r}}$ in $u_k$ and $v_k$ as
\begin{eqnarray}
u_{\bf k}({\bf r})={1 \over \sqrt{V}}e^{ i {\bf k} \cdot {\bf r}}u_k, \quad
v_{\bf k}({\bf r})={1 \over \sqrt{V}}e^{i {\bf k} \cdot {\bf r}}v_k
\end{eqnarray}
the field operators become
\begin{eqnarray}
\hat{\Psi}_{\uparrow}({\bf r})&=&\sum_{\bf k} e^{{i \over 2}\hat{\varphi}({\bf r})} \left( \gamma_{{\bf k} \uparrow } u_{\bf k}({\bf r})  -\gamma^{\dagger}_{{\bf k} \downarrow } v^{\ast}_{\bf k}({\bf r}) \right)
\nonumber
\\
\hat{\Psi}_{\downarrow}({\bf r})&=&
\sum_{\bf k} e^{{i \over 2}\hat{\varphi}({\bf r})}\left( \gamma_{{\bf k} \downarrow } u_{\bf k}({\bf r}) +\gamma^{\dagger}_{{\bf k} \uparrow } v^{\ast}_{\bf k}({\bf r}) \right)
\end{eqnarray}
where the spatial dependence is included in $\hat{\varphi}$.

 Now we allow the coordinate dependent functions that are different from plane waves. 
 We use the label $n$ in place of the wave number ${\bf k}$. Then, the field operators become
\begin{eqnarray}
\hat{\Psi}_{\uparrow}({\bf r})&=&\sum_{n} e^{{i \over 2}\hat{\varphi} ({\bf r})}\left( \gamma_{{n} \uparrow } u_{n}({\bf r})  -\gamma^{\dagger}_{{n} \downarrow } v^{\ast}_{n}({\bf r}) \right)
\nonumber
\\
\hat{\Psi}_{\downarrow}({\bf r})&=&
\sum_{n} e^{{i \over 2}\hat{\varphi} ({\bf r})} \left( \gamma_{{n} \downarrow } u_{n}({\bf r}) +\gamma^{\dagger}_{{n} \uparrow } v^{\ast}_{n}({\bf r}) \right)
\end{eqnarray}

The particle number conserving Bogoliubov operators satisfy
\begin{eqnarray}
\gamma_{n \sigma}|{\rm Gnd}(N) \rangle=0, \quad  \langle {\rm Gnd}(N)| \gamma^{\dagger}_{n \sigma}=0
\end{eqnarray}
where $N$ is the total number of particles.

We assume the ground state to be the eigenstate of $e^{{i \over 2} \hat{\varphi}({\bf r})}$. Then, it satisfies 
\begin{eqnarray}
e^{{i \over 2} \hat{\varphi}({\bf r})}|{\rm Gnd}(N) \rangle= e^{{i \over 2} {\varphi}({\bf r})}|{\rm Gnd}(N-1) \rangle
\label{eqBC}
\end{eqnarray}

The electronic Hamiltonian is expressed using the field operators as
\begin{eqnarray}
H=\sum_{\sigma} \int d^3 r \hat{\Psi}^{\dagger}_{\sigma}({\bf r}) h({\bf r}) \hat{\Psi}_{\sigma}({\bf r}) 
-{1 \over 2} \sum_{\sigma, \sigma'}\int d^3 r d^3 r' V_{\rm eff}({\bf r}, {\bf r}') \hat{\Psi}^{\dagger}_{\sigma}({\bf r}) \hat{\Psi}^{\dagger}_{\sigma'}({\bf r}') \hat{\Psi}_{\sigma'}({\bf r}') \hat{\Psi}_{\sigma}({\bf r}) 
\end{eqnarray}
where $h({\bf r})$ is the single-particle Hamiltonian given by
\begin{eqnarray}
h({\bf r})={ 1 \over {2m_e}} \left( { \hbar \over i} \nabla +{e \over c} {\bf A}^{\rm em} \right)^2+U({\bf r})-\mu 
\end{eqnarray}
and $-V_{\rm eff}$ is the effective interaction between electrons.

We perform the mean field approximation 
\begin{eqnarray}
H^{\rm MF}&=&\sum_{\sigma} \int d^3 r \hat{\Psi}^{\dagger}_{\sigma}({\bf r}) h({\bf r}) \hat{\Psi}_{\sigma}({\bf r}) 
+\int d^3 r d^3 r' 
\left[ \Delta({\bf r}, {\bf r}')\hat{\Psi}^{\dagger}_{\uparrow}({\bf r}) \hat{\Psi}^{\dagger}_{\downarrow}({\bf r}') e^{{i \over 2}(\hat{\varphi}({\bf r}) +\hat{\varphi}({\bf r}')) }
+{\rm H. c.} \right]
\nonumber
\\
&+&\int d^3 r d^3 r' 
{ {|\Delta({\bf r}, {\bf r}')|^2} \over {V_{\rm eff}({\bf r}, {\bf r}') }}
\end{eqnarray}
where the gap function $\Delta({\bf r}, {\bf r}')$ is defined as 
\begin{eqnarray}
 \Delta({\bf r}, {\bf r}')= V_{\rm eff}({\bf r}, {\bf r}')\langle e^{-{i \over 2}(\hat{\varphi}({\bf r}) +\hat{\varphi}({\bf r}')) }
\hat{\Psi}_{\uparrow}({\bf r}) \hat{\Psi}_{\downarrow} ({\bf r'}) \rangle
\end{eqnarray}

Due to the factor $ e^{-{i \over 2}(\hat{\varphi}({\bf r}) +\hat{\varphi}({\bf r}'))}$ the expectation value can be calculated using the particle number fixed state.

Using commutation relations for $\hat{\Psi}^{\dagger}_{\sigma }({\bf r})$ and $\hat{\Psi}_{\sigma }({\bf r})$,
\begin{eqnarray}
&&\{ \hat{\Psi}_{\sigma }({\bf r}),\hat{\Psi}^{\dagger}_{\sigma' }({\bf r}') \}=\delta_{\sigma \sigma'}\delta({\bf r} -{\bf r}')
\nonumber
\\
&&\{ \hat{\Psi}_{\sigma }({\bf r}),\hat{\Psi}_{\sigma' }({\bf r}') \}=0
\nonumber
\\
&&\{ \hat{\Psi}^{\dagger}_{\sigma }({\bf r}),\hat{\Psi}^{\dagger}_{\sigma' }({\bf r}') \}=0
 \end{eqnarray}
the following relations are obtained
\begin{eqnarray}
\left[\hat{\Psi}_{\uparrow }({\bf r}) , {H}_{\rm MF} \right]&=&
{h}({\bf r})\hat{\Psi}_{\uparrow }({\bf r})+\int d^3 r' \Delta({\bf r},{\bf r}')\hat{\Psi}^{\dagger}_{\downarrow }({\bf r}')e^{{i \over 2}(\hat{\varphi}({\bf r}) +\hat{\varphi}({\bf r}')) }
\nonumber
\\
\left[\hat{\Psi}_{\downarrow }({\bf r}) , {H}_{\rm MF} \right] &=&{h}({\bf r})\hat{\Psi}_{\downarrow }({\bf r})-\int d^3 r' \Delta({\bf r},{\bf r}')\hat{\Psi}^{\dagger}_{\uparrow }({\bf r}')e^{{i \over 2}(\hat{\varphi}({\bf r}) +\hat{\varphi}({\bf r}')) }
\label{deG1}
\end{eqnarray}

The particle number conserving Bogoliubov operators $\gamma_{n \sigma}$ and $\gamma^{\dagger}_{n \sigma}$ obey fermion commutation relations. They are chosen to satisfy
\begin{eqnarray}
\left[ {H}_{\rm MF}, \gamma_{n \sigma } \right]=-\epsilon_n \gamma_{n \sigma}, \quad \left[{H}_{\rm MF}, \gamma^{\dagger}_{n \sigma } \right] =\epsilon_n \gamma^{\dagger}_{n \sigma}
\label{deG2}
\end{eqnarray}
with $\epsilon_n \geq 0$. Then, ${H}_{\rm MF}$ is diagonalized as
\begin{eqnarray}
{H}_{\rm MF}=E_g + \sum_{n, \sigma} \epsilon_n \gamma^{\dagger}_{n \sigma}\gamma_{n \sigma}
\label{deG3}
\end{eqnarray}
where $E_g$ is the ground state energy.

From Eqs.~(\ref{deG1}), (\ref{deG2}), and (\ref{deG3}), we obtain the following system of equations
\begin{eqnarray}
\epsilon_n e^{{i \over 2} \hat{\varphi}({\bf r})}u_n({\bf r})&=&
h({\bf r}) e^{{i \over 2}\hat{\varphi}({\bf r})}u_n({\bf r})+\int d^3 r' \Delta ({\bf r},{\bf r}')e^{{i \over 2}\hat{\varphi}({\bf r})}v_n({\bf r}')
\nonumber
\\
\epsilon_n e^{{i \over 2} \hat{\varphi}({\bf r})}v^{\ast}_n({\bf r})&=&-
 h({\bf r}) e^{{i \over 2} \hat{\varphi}({\bf r})}v^{\ast}_n({\bf r})+\int d^3r' \Delta ({\bf r},{\bf r}')e^{{i \over 2}\hat{\varphi}({\bf r})}u^{\ast}_n({\bf r}')
\end{eqnarray}

Using the relation in Eq.~(\ref{eqBC}),
the above are cast into the following,
\begin{eqnarray}
\epsilon_n u_n({\bf r})&=&
\bar{h}({\bf r}) u_n({\bf r})+\int d^3 r'\Delta ({\bf r},{\bf r}')v_n({\bf r}')
\nonumber
\\
\epsilon_n v_n({\bf r})&=&-
 \bar{h}^{\ast}({\bf r}) v_n({\bf r})+\int d^3 r'\Delta^{\ast}({\bf r},{\bf r}')u_n({\bf r}')
 \label{e1}
\end{eqnarray}
where 
\begin{eqnarray}
\bar{h}({\bf r})={ 1 \over {2m_e}} \left( { \hbar \over i} \nabla +{e \over c} {\bf A}^{\rm em}+{ \hbar \over 2} \nabla \varphi \right)^2+U({\bf r})-\mu 
 \label{e2}
\end{eqnarray}
and 
\begin{eqnarray}
\Delta({\bf r}, {\bf r}')=V_{\rm eff}({\bf r}, {\bf r}')\sum_n \left[ u_n({\bf r}) v^{\ast}_n({\bf r}')(1- f(\epsilon_n))-u_n({\bf r}') v^{\ast}_n({\bf r})f(\epsilon_n) \right]
 \label{e3}
\end{eqnarray}
$f(\epsilon_n)$ is the Fermi function. They are Bogoliubov-de~Gennes equations \cite{deGennes} 
using the particle number conserving Bogoliubov operators \cite{koizumi2019}.
The gauge potential in the single particle Hamiltonian $\bar{h}({\bf r})$ is the effective one given by
\begin{eqnarray}
{\bf A}^{\rm eff}={\bf A}^{\rm em}+{ {\hbar c} \over {2e}} \nabla \varphi
\end{eqnarray}
 
If we solve the system of equations composed of Eqs.~(\ref{e1}), (\ref{e2}), and (\ref{e3}), with the condition ${\bf A}^{\rm em}+{ {\hbar c}\over {2e}} \nabla \varphi=0$, we obtain the currentless solutions
for $u_n, v_n$, which we denote as $\tilde{u}_n, \tilde{v}_n$. 

When a magnetic field is present, we may construct the solution using $\tilde{u}_n, \tilde{v}_n$ 
as
\begin{eqnarray}
u_n({\bf r})=\tilde{u}_n ({\bf r}) e^{{i \over 2}{\varphi}({\bf r})}, \quad v_n({\bf r})=\tilde{v}_n ({\bf r}) e^{-{i \over 2}{\varphi}({\bf r})}
\label{equv}
\end{eqnarray}
with suitably chosen $\nabla \varphi$.
In this case, we have
\begin{eqnarray}
m_e{\bf v}={ e\over c}{\bf A}^{\rm em}+{\hbar \over 2} \nabla \varphi
\label{Aeff}
\end{eqnarray}
since the velocity field from $\tilde{u}_n, \tilde{v}_n$ is zero.

Comparison of this with Eq.~(\ref{eqr1}) yields 
\begin{eqnarray}
m=m_e
\end{eqnarray}
This shows that the mass in the London moment is the free electron mass.

From Eq.~(\ref{equv}), we also have
\begin{eqnarray}
\nabla \chi^{\rm super}={1 \over 2} \nabla \varphi
\end{eqnarray}
Thus, we can identify the Berry phase from many-body wave functions as the London's superpotential.

London envisaged that the superconductivity is characterized by the presence of the long-range average momentum produced by the superpotential;
this is rephrased in the present theory that the superconductivity is characterized by the presence of the long-range Berry connection.

 \section{Relation to the Ginzburg-Landau theory}
 \label{sec5}
 
 Let us consider the case where $V_{\rm eff}$ is given by
 \begin{eqnarray}
 V_{\rm eff}({\bf r}, {\bf r}')=V_{\rm eff}({\bf r})\delta({\bf r}-{\bf r}')
 \end{eqnarray}
 
 In this case, the energy gap is given by
 \begin{eqnarray}
\Delta({\bf r})=V_{\rm eff}({\bf r})\sum_n u_n({\bf r}) v^{\ast}_n({\bf r}) \tanh {{ \epsilon_n} \over {2 k_{\rm B} T}}
\label{eqDelta1}
\end{eqnarray}

When the magnetic field is absent, it is given by
 \begin{eqnarray}
\tilde{\Delta}({\bf r})=V_{\rm eff}({\bf r})\sum_n \tilde{u}_n({\bf r}) \tilde{v}^{\ast}_n({\bf r}) \tanh {{\epsilon_n} \over {2 k_{\rm B} T}}
\label{eqDelta2}
\end{eqnarray}

From Eqs.~(\ref{equv}), (\ref{eqDelta1}), and (\ref{eqDelta2}), the energy gap in the presence of the magnetic field is given by
 \begin{eqnarray}
 {\Delta}({\bf r})=\tilde{\Delta}({\bf r})e^{{i }{\varphi}({\bf r})}
 \end{eqnarray}
 
 The macroscopic wave function in the Ginzburg-Landau theory is proportional to the energy gap in the standard theory.
 The momentum operator for it in the magnetic field should be given by the following change
  \begin{eqnarray}
 { \hbar \over i} \nabla \rightarrow  { \hbar \over i} \nabla +{{2e} \over c} {\bf A}^{\rm em}
 \end{eqnarray}
 to have the combination of ${\bf A}^{\rm em}$ and $\nabla \varphi$ given in the right-hand side of Eq.~(\ref{Aeff}).
 
 This above change is the one found in the Ginzburg-Landau theory. 
Since $\tilde{u}_n$ and $\tilde{v}_n$ can be obtained by the standard theory, the free energy for the Ginzburg-Landau theory is unchanged 
in the present theory.

 \section{Relation of the collective mode arising from the Berry connection and the Nambu-Goldstone mode}
 \label{sec6}
 
 The BCS theory has explained the Meissner effect by calculating the induced current as a linear response to the vector potential \cite{BCS1957}.
 The gauge invariance of the induced current was questioned since the BCS calculation was based on the particular gauge choice for the vector potential, and if other gauge had been employed a different result would have been obtained.
 
 This gauge invariance problem was solved by Nambu by taking into account the corrections arising from the Ward-Takahashi identity \cite{Ward,Takahashi57}. Thereby, the appearance of the longitudinal mode called the {\em Nambu-Goldstone} was discovered \cite{Nambu1960}.
  
The appearance of this mode, however, requires the breakdown of the global $U(1)$ invariance.
 It has been claimed by a number of researchers that such a breakdown is impossible since the relevant Hamiltonian for superconductivity conserves the particle number \cite{WWW1970,Peierls1991,Peierls92,LeggettBook}. Even if a pure state of a linear combination of different particle number states is prepared, it will soon be collapsed to a mixed state of different particle number states by the decoherence caused by the interaction with environment \cite{Zurek2002}. 
 
 Actually, the Meissner effect means not merely the exclusion of the magnetic field; it also means the superconducting state is a thermodynamically stable state in the $T$-$H$ plane \cite{London1950}. 
 Then, the supercurrent should be calculated by the formula given in Eq.~(\ref{eqMeisnner1}) \cite{Schafroth}. 
 Indeed, the Ginzburg-Landau theory uses this method.
 
 The use of the linear response theory to calculate the supercurrent is problematic since it is tided to the fluctuation-dissipation relation \cite{Kubo1957}; the current calculated by it usually causes dissipation, which contradicts the reversible superconducting-normal phase transition observed in the type I superconductors.
 It is also noteworthy that if the perturbation used in the linear response is derived from some kind of force in classical mechanics,
the mass appears there should be the effective mass, the inertial mass for the force in the solid.

We regard the induced current calculated by the linear response theory as something corresponding to the linear approximation of the current obtained by Eq.~(\ref{eqMeisnner1}), given in Eq.~(\ref{eqMeisnner2}), and examine the relation between the collective mode arising from the Berry connection and the Nambu-Goldstone mode, below.

We can express the current in Eq.~(\ref{eqMeisnner1}) as follows,
\begin{eqnarray}
{\bf j}({\bf r})=- {{2e} \over {\hbar}} 
{{\delta F[{\bf A}^{\rm em} + {{c \hbar} \over {2e}} \nabla \varphi
] } \over {\delta \nabla \varphi({\bf r})} }
\label{eqMeisnner3}
\end{eqnarray}

This indicates that the $\nabla \varphi$ cannot be obtained by merely the minimization of the free energy
since the obtained $\nabla \varphi$ by the minimization will satisfy 
\begin{eqnarray}
{{\delta F[{\bf A}^{\rm em} + {{c \hbar} \over {2e}} \nabla \varphi
] } \over {\delta \nabla \varphi({\bf r})} }=0
\end{eqnarray}

 This is in accordance with the so-called `Bloch's theorem' \cite{Bohm-Bloch,Bloch1966}.
Actually, the determination of  $\nabla \varphi$ has to be done by under constrains. Then, ${\bf j}({\bf r})\neq0$ becomes possible.

The first constraint is the conservation of charge \cite{koizumi2019,koizumi2020c}.
Nambu utilized the Ward-Takahashi identity to obtain the Nambu-Goldstone mode, and the use of the Ward-Takahashi identity is equivalent to 
imposing the conservation of charge. 
Therefore, the appearance of the Nambu-Goldstone mode in the standard theory may be interpreted that
something corresponding to the London's superpotential is needed to have the conservation of the charge in the original BCS theory \cite{BCS1957}.

It is also notable that in the process of minimizing $F$, the arbitrariness of the gauge in ${\bf A}^{\rm em}$ is absorbed by ${\bf A}^{\rm fic}$;
thus, the gauge invariance problem in ${\bf j}$ does not arise if the theory is equipped with the superpotential.

Nambu also assumed that the Bogoliubov transformation that mixes different particle number states is valid \cite{Nambu1960}; thereby 
the ground state with the broken global $U(1)$ invariance state is obtained.
Actually, the Bogoliubov type excitations can be expressed using the particle number conserving operator as shown in Section~\ref{sec4}. 
The number changing operators there express the change of number of particles participating in the collective motion, and the Bogoliubov type excitations are those associated with the transfer of the particles between the collective mode and the single-particle modes.
Thus, the change of the total particle number assumed in the standard theory is replaced by the change of the particle number in the collective mode in the present theory. 
Then, the breakdown of the global $U(1)$ gauge invariance is discarded from the theory of superconductivity.
 
Berry required the single-valuedness of the parameter dependent wave function with respect to the parameter \cite{Berry}. In the present case, the parameter is the particle coordinate, thus, the determination of $\nabla \varphi$ also requires the single-valuedness of wave functions with respect to the particle coordinates. It is also noteworthy that Schr\"{o}dinger imposed the single-valuedness of the wave function with respect to the coordinate as one of the postulates of quantum mechanics \cite{Schrodinger}.

This single-valuedness condition gives rise to the flux quantization. The Meissner effect yields ${\bf j}=0$ deep inside the supercoductor, and Eq.~(\ref{eqMeisnner2}) means for this case,
\begin{eqnarray}
{\bf A}^{\rm eff}=0
\end{eqnarray}

Then, by integrating the above along a closed loop $C$ deep inside the superconductor, we have
\begin{eqnarray}
\oint_C{\bf A}^{\rm em} \cdot d{\bf r}=-\oint_C{\bf A}^{\rm fic}\cdot d{\bf r}=-{{ c \hbar } \over {2e}} \oint_C \nabla \varphi \cdot d{\bf r}
\end{eqnarray}
The single-valuedness of the wave function requires 
\begin{eqnarray}
\oint_C \nabla \varphi \cdot d{\bf r} = 2 \pi n, \quad \mbox{ $n$ is a integer}
\end{eqnarray}
giving the flux quantization,
\begin{eqnarray}
\oint_C{\bf A}^{\rm em} \cdot d{\bf r}=-{{ c h } \over {2e}}n
\end{eqnarray}

The number `$2$' appearing here is due to the property of the Berry connection, not due to the electron pairing. 

London has shown that the current distribution and magnetic field in the superconductor are uniquely determined by the fluxes for $\nabla \chi^{\rm super}$ and the boundary magnetic field \cite{London1950}. Thus, the $\nabla \varphi$ can be uniquely determined by the requirement of the conservation of the charge and fluxes for $\nabla \varphi$ under the constraint of the single-valuedness of the wave function with respect to the particle coordinates. In other words, we can obtain $\nabla \varphi$ by the free energy minimization under the constraints of the charge conservation and the single-valuedness of the wave function. 
The supercurrent should be calculated using and calculate the supercurrent using Eq.~(\ref{eqMeisnner1}).

In the standard theory, the superconductivity is characterized by the broken gauge symmetry and the appearance of the Nambu-Goldstone mode \cite{Anderson,Weinberg};
this is rephrased in the present theory that the superconductivity is characterized by the presence of the non-trivial Berry connection (or the gauge field that make ${\bf A}^{\rm eff}$ gauge invariant) and the appearance of the accompanying collective mode.

Note that the origin of $\nabla \varphi$ is outside the BCS theory in the present formalism. A possible mechanism for the appearance of it has been given in our previous work \cite{koizumi2020}.

\section{Concluding remarks}
 \label{sec7}

In many-body systems, each particle is in the gauge field given by Eq.~(\ref{eqBerryConnection}). This is the gauge field arising from the interaction through the wave function they form.
It causes the replacement of the spatial derivative as shown in Eq.~(\ref{eqFBerry}) although its effect is not usually noticeable.

However, when a collective mode is created from it, its effect is spectacular. In this case, the system is equipped with the common connection, the Berry connection, of the geometry in mathematics.
It has been argued that the superfluid phenomena can be considered as due to the presence of this collective mode \cite{koizumi2019}.

We have argued that the non-trivial Berry connection from many-body wave functions plays the role of the long-range average momentum produced as the gradient of the superpotential, and the collective mode arising from it replaces the Nambu-Goldstone mode in the standard theory.
In this way, the mass in the London moment becomes the free electron mass in agreement with the experiments.
The breakdown of the global $U(1)$ invariance assumed in the standard theory is discarded, and the supercurrent becomes a thermodynamically stable one calculated using the free energy in agreement with the reversible superconducting-normal phase transitions in the magnetic field.

\acknowledgments{The author thanks Dr.~Andras Kovacs for the informing about the London moment problem.}
%\bibliography{SPIN-BCS}

\end{document}